\documentclass[aps,prb,showpacs,intlimits,amsmath,amssymb,twocolumn]{revtex4-1}
\usepackage{graphicx}
\usepackage{bm}

\begin{document}

\title{Time-domain pumping a quantum-critical charge-density-wave-ordered material}

\author{O.~P.~Matveev$^{1,2}$,  A.~M.~Shvaika$^2$,
 T.~P.~Devereaux$^{3,4}$, and J.~K.~Freericks$^1$}
\affiliation{$^1$ Department of Physics, Georgetown University, Washington, DC
20057, USA}
\affiliation{$^2$ Institute for Condensed Matter Physics of the National Academy of Sciences of Ukraine,
Lviv, 79011 Ukraine}
\affiliation{$^3$ Geballe Laboratory for Advanced Materials, Stanford University, 
Stanford, CA 94305,USA}
\affiliation{$^4$ Stanford Institute for Materials and Energy Sciences (SIMES), 
SLAC National Accelerator Laboratory, Menlo Park, CA 94025, USA }

\begin{abstract}

We determine the exact time-resolved photoemission spectroscopy for a nesting driven charge-density-wave (described by the spinless 
Falicov-Kimball model within dynamical mean-field theory).  The pump-probe experiment involves  
two light pulses: the first is an ultrashort intense pump pulse that excites the system into nonequilibrium, and 
the second is a lower amplitude higher frequency probe pulse that photoexcites electrons.  We examine three different cases: the strongly correlated metal, the quantum-critical charge density wave and the critical Mott insulator.  Our results show that the quantum critical charge density wave has an ultra efficient relaxation channel that
allows electrons to be de-excited during the pump pulse, resulting in little net excitation. In contrast, the metal and the Mott insulator show excitations that are closer to what one expects from these systems. In addition, the pump field produces spectral band narrowing, peak sharpening, and a spectral gap reduction,
all of which rapidly return to their field free values after the pump is over.

\end{abstract}

\maketitle

\section{Introduction}

Charge-density-wave (CDW) systems are interesting as they exhibit an order parameter given by the
modulation of the charge density of electrons in real space. Often the ordering pattern is commensurate with the lattice, which
means the translational invariance of the system is partially broken. In other cases, the order
is incommensurate with the lattice, which proves to be much more difficult to simulate. In electron-mediated
CDW's the ordering disappears by the build-up of subgap states which can open novel conducting channels. In particular, the CDW has a quantum critical point which is a metal-to-insulator critical point, being insulating at $T=0$ and metallic for nonzero $T$ due to the emergence of subgap states at the chemical potential precisely as $T$ increases from zero. We study how these subgap states affect the 
time-domain pumping of the CDW, especially with regards to photoemission experiments.

Recently,
there has been significant interest in pump-probe experiments on these materials including
photoemission spectroscopy (PES), core-level PES (XPS), and electron diffraction. In particular, $\text{TbTe}_3$,~\cite{schmitt1,schmitt2} $\text{TaS}_2$,~\cite{perfetti1,perfetti2,hellmann1,hellmann2,kirchmann} and 
$\text{TiSe}_2$,~\cite{hellmann2,rohwer} were investigated with pump-probe angle-resolved PES, which provides both time and angle resolution. One of the reasons for performing these experiments was to try to resolve whether the CDW order is mediated by the electrons, by the phonons, or by the electron-phonon coupling\cite{mazin}. Time-domain experiments have the potential to separate out these effects on short time scales---electron
mediated interactions should be fast, and phonon mediated ones slow on fs time scales. In this sense, TiSe has been identified as an electron-mediated CDW (and possibly an excitonic CDW\cite{hellmann2,rohwer,eckstein}). The nonequilibrium driving of these systems has also produced new ``nonequilibrium phases'' which do not occur in equilibrium~\cite{stojchevska,vaskivskyi,han}. 

In addition, the PES signal in the above experiments showed that 
the CDW gap generically closes for a short period of time, but at the same time the modulation of the charge density remains nonzero. 
The above experiments also displayed oscillations of the PES signal at long times after the pump pulse is gone, which oscillate at the frequency of the phonon responsible for the ordering; we do not model such behavior in the all-electron model studied here.
An initial theoretical study was conducted 
on the simplest noninteracting CDW system\cite{shen_fr1}. The PES response shows that the CDW gap closes (while the order parameter remains nonzero) when the pump is 
on and it restores after the pump pulse is turned off.

In order to address how a CDW behaves after being pumped, we calculate the PES response for the CDW ordered phase of the Falicov-Kimball model.\cite{falicov_kimball} This model possesses both a metal-insulator
phase transition and  a transition from uniform to commensurate or incommensurate CDW ordered phases in equilibrium.\cite{brandt_mielsch1,freericks_review} It has
an exact solution in nonequilibrium\cite{frturzlat_prl97,fr_prb77} within the dynamical mean-field theory (DMFT). The equilibrium density of states (DOS) displays
nontrivial behavior in the ordered phase: the  width of the CDW gap in the DOS (which occurs at zero temperature) does not change when the temperature increases, in opposition to what happens in BCS
theory\cite{bcs}, where the gap in the DOS continuously closes as $T_c$ is approached from below. Instead, it initially fills the gap region with subgap states which increase until the gap is fully filled at the critical temperature.\cite{hass_krishn,msf_opt,lemanski} Another feature of this model is that there is a critical value of the interaction where the subgap states start to form at the chemical potential just as $T$ increases above zero. This quantum critical state has an instantaneous transition from an insulator at $T=0$ to a metal for any finite temperature.
Hence, the Falicov-Kimball model shows additional correlation effects and our work goes beyond the previous calculations of the PES signal on the simplest 
noninteracting CDW model.\cite{shen_fr1} We expect our results to be useful in experiments on real CDW materials such as those already mentioned above.

The remainder of the paper is organized as follows: In Sec.~\ref{sec:1} we present the theory for our calculations. It consists of two subsections: In Subsec.~\ref{subsec:1}, 
we define the time-dependent Hamiltonian of system in the CDW ordered state. In Subsec.~\ref{subsec:2}, we describe the theory for the photoemission response 
function. In Sec.~\ref{sec:2}, we present our results for PES function in the CDW ordered phase and discussion. We conclude in Sec.~\ref{sec:3}.

\section{Formalism}
\label{sec:1}

We develop the nonequilibrium DMFT to solve for the two-time contour-ordered Green's function defined on the Kadanoff-Baym-Keldysh time contour. We generalize the theory to encompass CDW ordered phases of the Falicov-Kimball model which requires an additional $2\times 2$ matrix structure. Finally, we derive the formulas for the time-resolved PES response function.  

\subsection{CDW ordered state Hamiltonian}
\label{subsec:1}

The Falicov-Kimball model involves two kinds of particles: heavy electrons which are localized on the lattice sites, and light itinerant 
electrons~\cite{falicov_kimball} which are allowed to hop between nearest-neighbor sites. The model possesses a transition into a commensurate bipartite CDW 
ordered phase~\cite{brandt_mielsch1,freericks_review} at low temperature. At half-filling, this occurs for any value of the Coulomb interaction and sufficiently low $T$. 
Employing the Kadanoff-Baym-Keldysh approach\cite{kadanoff,keldysh}, the nonequilibrium DMFT formalism was developed previously for the uniform 
phase\cite{frturzlat_prl97,fr_prb77,frtur_prb71}. Recently, we generalized it for the case of the bipartite CDW ordered phase\cite{noneq_cdw_jsnm,noneq_cdw_prb}. Here, we only summarize  the main steps of the theory to establish our notation. 

The ordering arises from the nesting instability of the Fermi surface with a modulation wavevector given by $\mathbf{Q} =(\pi, \pi, \dots)$. We employ two sublattices 
``$A$'' and ``$B$'' which have different electron density (in a checkerboard pattern). This implies that the Brillouin zone is cut in half and is called the reduced Brillouin zone (rBZ). To define each
sublattice, we apply the modulation wavevector $\mathbf{Q}$ as follows:
\begin{align}\label{Q}
e^{i\mathbf{Q}\cdot\mathbf{R}_i}=\begin{cases}
                                 ~~1, & \quad \mathbf{R}_i\in A, \\
                                -1, & \quad \mathbf{R}_i\in B,
                            \end{cases} 
\end{align}
where ${\bf R}_i$ is the position vector for the $i$th lattice site. 

We express the Hamiltonian in terms of fermionic annihilation and creation operators with respect to this underlying two-sublattice system. We can do this 
either in  real space or in momentum space. In the real space picture, we simply add an extra index for the sublattice:
\begin{equation}
 c_i^{\phantom{\dagger}}\rightarrow c_{i,\alpha}^{\phantom{\dagger}},\quad \alpha=A,B.
 \label{cAB}
\end{equation}
Performing a Fourier transform to momentum space, but with the summation over the lattice restricted to be either over the $A$ sublattice or the $B$ sublattice only, produces the corresponding momentum-dependent
operators $c_{\mathbf{k},\alpha}^{\phantom\dagger}$.
In the momentum space picture, we introduce two new operators which define the fermionic operators for the $\bf k$ and $\bf{k}+\bf{Q}$ subspaces as follows:   
\begin{equation}
 \tilde{c}_{1\bf k}^{\phantom{\dagger}} = c_{\bf k}^{\phantom{\dagger}}\quad \text{and} \quad \tilde{c}_{2\bf k}^{\phantom{\dagger}} = c_{\bf{k+Q}}^{\phantom{\dagger}},
 \label{ckQ}
\end{equation}
where the momentum $\bf{k}$ is restricted to the rBZ and $\mathbf{k}+\mathbf{Q}$ is restricted to the complement of the rBZ. The same notation is employed for the creation operators.  
Applying a Fourier transformation, we find that the relation between the real space operators in Eq.~(\ref{cAB}) and the momentum space operators in Eq.~(\ref{ckQ}) 
can be written in matrix form:     
\begin{align}
\label{transformation}
   \begin{bmatrix}
  \tilde{c}_{1\mathbf k} \\
  \tilde{c}_{2\mathbf k}
   \end{bmatrix}
=\hat{U} \begin{bmatrix}
  c_{\mathbf k A} \\
  c_{\mathbf k B}
   \end{bmatrix}, \quad\text{where}\quad
\hat{U}= \begin{Vmatrix}
  \dfrac{1}{\sqrt{2}} &  \dfrac{1}{\sqrt{2}} \\
  \dfrac{1}{\sqrt{2}} & -\dfrac{1}{\sqrt{2}}
   \end{Vmatrix}.
\end{align}
The matrix $\hat{U}$ is unitary and the $\sqrt{2}$ factors are chosen to satisfy the standard commutation relations for the fermionic annihilation and creation 
operators.  The connection between any quantity that is constructed from two operators 
in the real space representation $\mathcal{\hat{O}}(\mathbf{k})$ (which is a $2\times2$ matrix in the ordered 
state), and in the momentum space representation $\mathcal{\hat{\widetilde{O}}}(\mathbf{k})$ follows from the unitary transformation via     
\begin{equation}
 \mathcal{\hat{\widetilde{O}}}(\mathbf{k})=\hat{U} \mathcal{\hat{O}}(\mathbf{k}) \hat{U}^{-1}.
\end{equation}

Employing this notation, the time-dependent Hamiltonian of the ordered system is written as
\begin{equation}\label{hamiltonian}
\mathcal{H}(t)=\sum_{i\alpha}\mathcal{H}^\alpha_i-\sum_{ij\alpha\beta} t^{\alpha\beta}_{ij}(t)c^{\dag}_{i\alpha} c_{j\beta}^{\phantom\dagger},
\end{equation}
where the local term describes the Coulomb interaction between the localized and itinerant electrons on the $i$th site of the $\alpha$th sublattice and 
the chemical potential (site energy):
\begin{equation}
\mathcal{H}^\alpha_i=U n^\alpha_{ic} n^\alpha_{if}-\mu n^\alpha_{ic}+E_f^\alpha n^\alpha_{if}.
\end{equation}
The number operators of the itinerant and localized electrons are given by 
$n^{\alpha}_{ic} = c_{i\alpha}^\dagger c_{i\alpha}^{\phantom\dagger}$ and $n_{if}^{\alpha}=f_{i\alpha}^\dagger f_{i\alpha}^{\phantom\dagger}$, respectively. 
The nonlocal kinetic-energy term of the Hamiltonian describes hopping of itinerant electrons between the nearest-neighbor sites (that belong to different 
sublattices---we will work on an infinite-dimensional hypercubic lattice). We work in units where $\hbar=c=e=a=1$.    

In a pump-probe experiment, the material is first irradiated with an intense ultrafast pump pulse to excite the electronic subsystem. Later, a higher frequency, lower amplitude probe
pulse is used to measure the temporal evolution of the nonquilibrium electrons. To model this scenario, we choose the pump pulse to be an electric field $\mathbf{E}(t)$ with a Gaussian envelope of the form   
\begin{equation}\label{efield}
\mathbf{E}(t)=\mathbf{E}_0\cos[\omega_p (t-t_0)]\exp{[-(t-t_0)^2/\sigma_p^2]},
\end{equation}
where $E_0=|\mathbf{E}_0|$ is the magnitude of the field at time $t=t_0$ (the maximum of the pump pulse). Here, we assume that the electric field is spatially uniform and that it is 
directed along the main diagonal in the infinite dimensional space (1,1,\ldots,1). We also ignore all magnetic field and relativistic effects. This allows us to 
describe the electric field via a spatially uniform vector potential in the Hamiltonian gauge:
\begin{equation}\label{field}
\mathbf{E}(t)=-\dfrac{d}{d t}\mathbf{A}(t).
\end{equation}
We exploit a Peierls' substitution to the kinetic-energy term of the Hamiltonian to describe the interaction between itinerant electrons and the external electric 
field in Eq.~(\ref{efield}). Hence, the hopping matrix depends on time explicitly as follows 
\begin{equation}\label{hopping}
t^{\alpha\beta}_{ij}(t)=t^{\alpha\beta}_{ij}\text{exp}\biggl(-i\int\limits_{\mathbf{R}_{i,\alpha}}^{\mathbf{R}_{j,\beta}}\mathbf{A}(t)\cdot d\mathbf{r}\biggr),
\end{equation}
where $t^{\alpha\beta}_{ij}$ is the (constant) hopping matrix in the absence of an external electric field.  
Performing a Fourier transformation to momentum space, we rewrite the time-dependent kinetic-energy term in the form  
\begin{align}
\hat{\mathcal{H}}_{kin}(t) &= \sum_{k}\begin{bmatrix}
c^{\dag}_{\mathbf{k}A} & c^{\dag}_{\mathbf{k}B}
\end{bmatrix}\hat{\epsilon}(\mathbf k-\mathbf{A}(t))\begin{bmatrix}
c_{\mathbf{k}A}^{\phantom\dagger}\\ c_{\mathbf{k}B}^{\phantom\dagger}
\end{bmatrix}
\nonumber
\\
&= \sum_{k}\begin{bmatrix}
\tilde{c}^{\dag}_{1\mathbf{k}} & \tilde{c}^{\dag}_{2\mathbf{k}}
\end{bmatrix}\hat{\tilde{\epsilon}}(\mathbf k-\mathbf{A}(t))\begin{bmatrix}
\tilde{c}_{1\mathbf{k}}^{\phantom\dagger}\\ \tilde{c}_{2\mathbf{k}}^{\phantom\dagger}
\end{bmatrix}.
\end{align}
Here, the extended band energy $\hat{\epsilon}(\mathbf k-\mathbf{A}(t))$\cite{frtur_prb71} is off-diagonal in the real space two-sublattice representation
\begin{align}\label{eps}
&\hat{\epsilon}(\mathbf k-\mathbf{A}(t)) \\
&=\begin{Vmatrix}
 \scriptstyle 0 & \scriptstyle\epsilon(\mathbf k)\cos (A(t))+\bar{\epsilon}(\mathbf k)\sin (A(t)) \\
 \scriptstyle \epsilon(\mathbf k)\cos (A(t))+\bar{\epsilon}(\mathbf k)\sin (A(t)) & \scriptstyle 0
 \end{Vmatrix},
\nonumber
\end{align}
where $\epsilon(\mathbf k)=-\lim_{d\rightarrow\infty}t^*\sum_{r=1}^d \cos k_r/\sqrt{d}$ and 
$\bar{\epsilon}(\mathbf k)=-\lim_{d\rightarrow\infty}t^*\sum_{r=1}^d \sin k_r/\sqrt{d}$ (we apply the same scaling of the hopping term 
as in equilibrium DMFT). The extended band energy in Eq.~(\ref{eps}) is diagonal in the momentum space representation.
By performing the unitary transformation in Eq.~(\ref{transformation}), we obtain for $\hat{\tilde{\epsilon}}(\mathbf k-\mathbf{A}(t))$
\begin{align}
\label{eps1}
&\hat{\tilde{\epsilon}}(\mathbf k-\mathbf{A}(t))
=\hat{U}\hat{\epsilon}(\mathbf k-\mathbf{A}(t))\hat{U}^{-1} \\
&=\begin{Vmatrix}
  \scriptstyle\epsilon(\mathbf k)\cos (A(t))+\bar{\epsilon}(\mathbf k)\sin (A(t)) & \scriptstyle 0 \\
  \scriptstyle 0 & \scriptstyle -\epsilon(\mathbf k)\cos (A(t))-\bar{\epsilon}(\mathbf k)\sin (A(t)) 
 \end{Vmatrix}.
\nonumber
\end{align}

The CDW ordered state is characterized by two order parameters. The heavy electron order parameter $\Delta n_f=(n_f^A-n_f^B)/2(n_f^A+n_f^B)$ is the 
difference of the heavy electron occupation on the $A$ and $B$ sublattices. It reaches its maximum value of $1/2$ at $T=0$ and becomes equal to $0$ 
at $T=T_c$. Since the heavy electrons do not interact with the external electric field, this order parameter does not change in time and remains fixed at its 
equilibrium value. While this may seem like an odd behavior, it arises because the heavy electrons arelocalized, and hence they remain fixed, even when a field is applied to the system. This behavior also
occurs in the simplified model where the CDW is determined by a bandstructure with a fixed checkerboard pattern to the site potential. In CDWs that arise due to a phonon distortion, the order parameter associated with that distortion can relax in time, but the behavior described above is what one would expect to see for short times.

The starting point for our calculations is an equilibrium state at a given temperature. To solve for the order parameter at this temperature, we avoid critical slowing down of the iterative DMFT process by working with a fixed order parameter, which determines the heavy electron filling on each sublattice~\cite{fk_cdw_prb03}. Then the heavy electron site energies $E_f^\alpha$ are calculated from the DMFT solution. The order parameter is adjusted until these two energies are equal, signifying the thermodynamic equilibrium state. At this point, we can determine the order parameter of the conduction electrons (which never reaches its maximal value of 1/2 due to Pauli blocking). It is given by the difference of the itinerant electron filling on the two sublattices $\Delta n_c(t)=[n_c^B(t)-n_c^A(t)]/2[n_c^A(t)+n_c^B(t)]$. Since itinerant electrons interact with the 
external electric field, this order parameter changes in time when the field is applied. Indeed, it is even distinct from the heavy electron order parameter when the
system is in equilibrium and it can change sign when the external pulse excites the system\cite{shen_fr1}. 

\subsection{Time-resolved PES response function}
\label{subsec:2}

\begin{figure} 
 \centerline{\includegraphics[height=0.12\textheight]{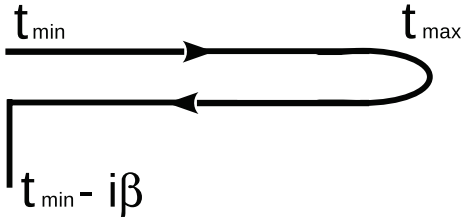}}
 \caption{Kadanoff-Baym-Keldysh time contour, which runs from a minimum time to a maximum time along the real time axis, then backwards to the minimum time, and then 
 parallel to the imaginary time axis for a length given by the inverse of the initial equilibrium temperature.}
 \label{contour}
\end{figure}

We exploit the theory for time-resolved, pump-probe, PES developed recently for the normal phase~\cite{pump-probe_theory}. This theory also holds in the CDW ordered 
phase, but in this case we have to generalize it for the two-sublattice system. The PES response function is computed from the lesser Green's function, which can be 
extracted from the contour-ordered Green's function. Since the Hamiltonian of the system depends on time explicitly, we have to apply the Kadanoff-Baym-Keldysh 
formalism to solve for the contour-ordered Green's function that depends on two times. In this case, the contour-ordered Green's function is defined on the 
Kadanoff-Baym-Keldysh time contour in Fig.~\ref{contour} as follows      
\begin{equation}\label{gf}
G^{c}_{\bf k}(t,t')=-i\langle\mathcal{T}_{c} c_{\bf k}^{\phantom\dagger}(t)c_{\bf k}^{\dag}(t')\rangle,
\end{equation}
where the average $\langle\mathcal{O}(t)\rangle = {\rm Tr} \exp[-\beta\mathcal{H}(t\rightarrow -\infty)]\mathcal{O}(t)/\mathcal{Z}$ is calculated in equilibrium, before the system is hit by the electric pulse, and we assume the Hamiltonian becomes time independent at early times. The partition function is 
$\mathcal{Z}={\rm Tr}\exp[-\beta\mathcal{H}(t\rightarrow -\infty)]$, and $\beta=1/T$ is the inverse of the initial equilibrium temperature. The continuous matrix operators must be converted to discrete matrices. This is done by employing  a finite discretization to the contour in Fig.~\ref{contour} with a fixed spacing $\Delta t=(t_{max}-t_{min})/N$ on its real branch and $\Delta\tau=\beta/n$ 
on its imaginary branch, so the Green's function in Eq.(\ref{gf}) is a $(2N+n)\times(2N+n)$ matrix~\cite{fr_prb77}. Further, for the set of 
$\{\Delta t_1>\Delta t_2>\Delta t_3>...\}$ ($\{N_1<N_2<N_3<...\}$), we 
extrapolate the results with a quadratic Lagrange interpolation formula to the zero spacing limit $\Delta t\rightarrow0$ ($N\rightarrow\infty$).

The lattice Green's function is a $2\times2$ block matrix in the ordered phase
\begin{equation}
\hat{G}^c_{\epsilon,\bar{\epsilon}}(t,t')=
\begin{Vmatrix}
 G^{c,AA}_{\epsilon,\bar{\epsilon}}(t,t') & G^{c,AB}_{\epsilon,\bar{\epsilon}}(t,t') \\[1em]
 G^{c,BA}_{\epsilon,\bar{\epsilon}}(t,t') & G^{c,BB}_{\epsilon,\bar{\epsilon}}(t,t')
\end{Vmatrix},
\label{eq: g_ab}
\end{equation}
where all the dependence on $\bf k$ is summarized by the two band energies [when the field is in the diagonal direction as in Eq.~(\ref{eps1})]; we adopt the following notation for the 
momentum-dependent Green's function: $G^{c}_{\bf k}(t,t')=G^c_{\epsilon,\bar{\epsilon}}(t,t')$.

We employ the standard iterative algorithm of nonequilibrium DMFT for the Falicov-Kimball model to determine the Green's functions~\cite{noneq_cdw_prb}. To verify our numerical results, we check the spectral moment sum rules which continue to hold  in the ordered phase and 
in nonequilibrium. Here we work with the local Green's functions which involve the two-dimensional integration of Eq.~(\ref{eq: g_ab}) over $\epsilon$ and $\bar\epsilon$ weighted by the joint density of states on the infinite-dimensional hypercubic latice; the local Green's functions are the diagonal elements of the $2\times 2$ matrix with the retarded Green's function extracted from the contour-ordered Green's function. The moments of the local retarded Green's function are defined as follows
\begin{align}\label{sumrules}
\mu_{n}^{R,a}(t_{a}) 
=-\dfrac{1}{\pi}\int\limits_{-\infty}^{\infty}d\omega\text{Im}\int\limits_{-\infty}^{\infty}dt_{r} e^{i\omega t_{r}}i^{n}\dfrac{\partial^n}{\partial t_r^n}G^{R,\alpha}(t_{a},t_{r}),
\end{align}
where $\alpha=(A,B)$ and we use the Wigner coordinates  for the average $t_{a}=(t+t')/2$ and relative $t_{r}=t-t'$ times. The local density of states on the $\alpha$ sublattice 
$A^{\alpha}(t_{a},\omega)=-\text{Im}G^{R,\alpha}(t_{a},\omega)/\pi$ is the Fourier transform of the retarded Green's function with respect to the relative 
time $t_{r}$. We calculate the zeroth, first and second moments which satisfy\cite{frtur_sumrules} 
\begin{equation}
 \mu_{0}^{R,\alpha}(T)=1, 
\end{equation}
\begin{equation}
 \mu_{1}^{R,\alpha}(T)=-\mu+Un_{f}^{\alpha}, 
\end{equation}
\begin{equation}
 \mu_{2}^{R,\alpha}(T)=\dfrac{1}{2}+\mu^{2}-2U\mu n_{f}^{\alpha}+U^{2}n_{f}^{\alpha},
\end{equation}
with $n_{f}^{A}=1/2+\Delta n_{f}$ and $n_{f}^{B}=1/2-\Delta n_{f}$. When we run calculations, we find that when the maximal pulse amplitude is large ($E_0=30$), we can verify that the extrapolated Green's functions have accurate spectral moments. On the other hand, when the field amplitude is small ($E_0=1$), then the accuracy is too poor to have trustworthy results unless the time steps are made prohibitively small\cite{spie}.

In the case of the CDW ordered phase, we calculate the time-resolved PES response function $P_{\alpha}(\omega,t_{0}{'})$ for each sublattice. It is double-time 
Fourier transform of the lesser Green's function weighted by the probe pulse envelope function $s(t)$ as follows~\cite{pump-probe_theory}
\begin{equation}
 P_{\alpha}(\omega,t_{0}{'})=-i\int\limits_{t_{\text{min}}}^{t_{{\text{max}}}}dt\int\limits_{t_{\text{min}}}^{t_{\text{max}}}dt' s(t) s(t') e^{-i\omega(t-t')}G^{<}_{\alpha}(t,t'),
\end{equation}
where $\alpha=A,B$. We assume the envelope function is a Gaussian of the form 
\begin{equation}\label{probe}
 s(t)=\dfrac{1}{\sigma_b \sqrt{\pi}} e^{-(t-t_{0}{'})^2/\sigma_b^2},
\end{equation}
where $t_{0}{'}$ is the time when the probe pulse has its maximum and it defines the time delay relative to the pump pulse maximum at $t_0$ in Eq.~(\ref{efield}) 
and $\sigma_b$ defines the effective width of the probe pulse. The width of the pulse determines the energy or time resolution of the PES response function: the broader width of the pulse, the better the energy resolution and the worse the time resolution, and \textit{vice versa} if it is narrower\cite{spie}. 
Because we work with the total PES and not the angle-resolved PES in this paper, we do not need to 
worry about gauge invariance. The PES response function is always manifestly gauge invariant.

\section{Results}
\label{sec:2}

\begin{figure}[htb]
\centerline{\includegraphics[width=0.5\textwidth]{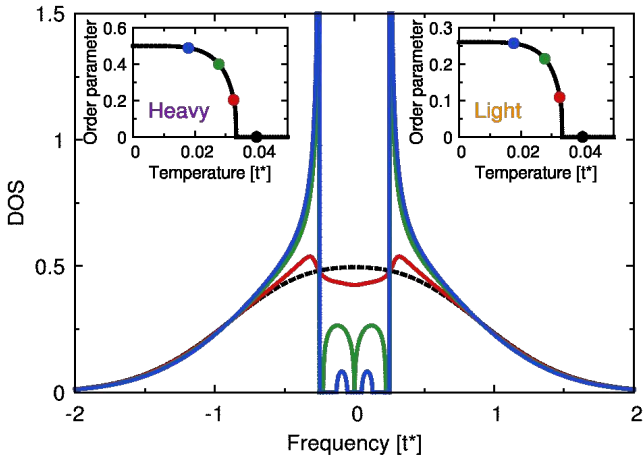}}
 \caption{(Color online.) Equilibrium DOS for $U=0.5$ (strongly correlated metal) at different temperatures: $T=0.0178$ corresponds to $\Delta n_f=0.49$ (blue); $T=0.0278$ 
 corresponds to $\Delta n_f=0.4$ (green); $T=0.0326$ corresponds to $\Delta n_f=0.2$ (red); $T=0.04$ corresponds to $\Delta n_f=0$ (black, dashed). 
 Insets show the temperature dependence of the corresponding order parameters of the localized (left) and itinerant (right) electrons. Note how the subgap DOS rapidly forms and fills in the gap region, while the signature of the spectral gap in the DOS remains fixed at $U$ for all $T$ in the ordered phase.}
 \label{dosu05}
\end{figure}

\begin{figure}[htb]
\centerline{\includegraphics[width=0.5\textwidth]{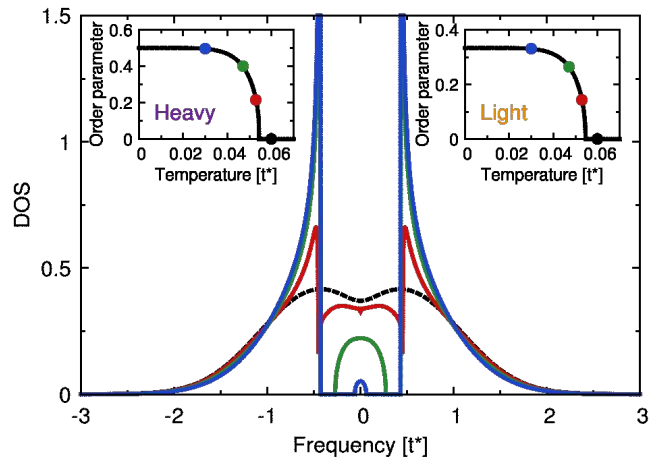}}
 \caption{(Color online.) Equilibrium DOS for $U=0.86$ (material at the CDW quantum critical point) at different temperatures: $T=0.03$ corresponds to $\Delta n_f=0.495$ (blue) ; $T=0.047$ 
 corresponds to $\Delta n_f=0.4$ (green); $T=0.053$ corresponds to $\Delta n_f=0.2$ (red); $T=0.06$ corresponds to $\Delta n_f=0$ (black dashed). 
 Insets show the temperature dependence of the order parameters of localized (left) and itinerant (right) electrons. Note how the subgap DOS rapidly forms at the chemical potential producing a metal and then fills in the gap region, while the signature of the gap in the DOS remains fixed at $U$ for all $T$ in the ordered phase.}
 \label{dosu086}
\end{figure}

\begin{figure}[htb]
\centerline{\includegraphics[width=0.5\textwidth]{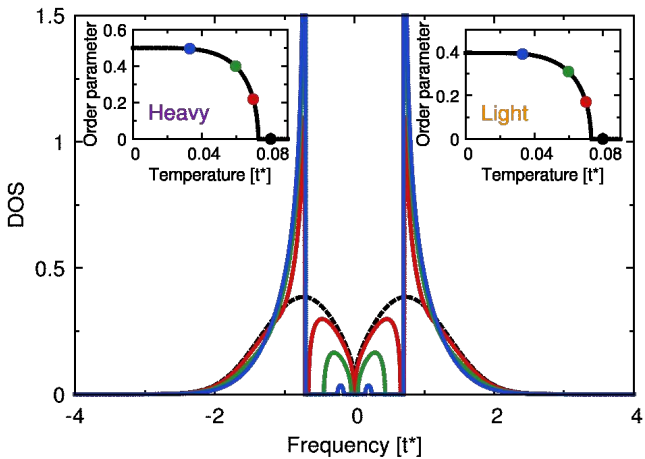}}
 \caption{(Color online.) Equilibrium DOS for $U=1.4$ (critical point for the Mott transition) at different temperatures: $T=0.033$ corresponds to $\Delta n_f=0.495$ (blue); $T=0.0596$ 
 corresponds to $\Delta n_f=0.4$ (green); $T=0.07$ corresponds to $\Delta n_f=0.2$ (red); $T=0.08$ corresponds to $\Delta n_f=0$ (black dashed). 
 Insets show the temperature dependence of the order parameters of the localized (left) and itinerant (right) electrons. Note how the subgap DOS rapidly forms but does not fill the entire gap region due to Mott physics suppressing the DOS at the chemical potential;  the signature of the gap in the DOS still remains fixed at $U$ for all $T$ in the ordered phase.}
 \label{dosu14}
\end{figure}

We present our results for the time-resolved pump-probe PES response function in the CDW ordered phase of the Falicov-Kimball model. Before discussing the 
nonequilibrium results, we show the equilibrium DOS's for three different Coulomb interactions $U$ in Figs.~\ref{dosu05}-\ref{dosu14}.

The temperature dependence of the equilibrium DOS is similar for all $U$ and it behaves as follows: at $T=0$, the equilibrium DOS shows a full CDW spectral gap whose width is precisely equal to the interaction $U$. In this case, the system is completely ordered: one sublattice is occupied by the $f$-electrons and the other sublattice is empty. This case corresponds precisely to the simplified CDW studied earlier~\cite{shen_fr2}. Increasing the 
temperature reduces the ordering and subgap states start to appear within the gap region\cite{msf_opt}.
These subgap states increase in magnitude, while the singularity at the gap edge is reduced (but maintains the same width $U$) until the CDW gap becomes completely closed at the critical temperature, 
and the order parameters vanish. Note, however, that the filling in of the subgap DOS always approaches that of the normal state, so if $U$ is large enough to be in the Mott insulating phase, no subgap states form within the Mott gap region (which is always smaller than $U$). We plot the order parameters 
$\Delta n_f$ and $\Delta n_c$ in the insets of Figs.~\ref{dosu05}-\ref{dosu14}.  

In nonequilibrium, when the system is pumped by the external 
electric field, the transient DOS shows significant changes in the gap region even for systems starting from zero temperature\cite{shen_fr1,shen_fr2}. At the same time, the itinerant electron order parameter has significant time dependence and often remains nonzero at the time when the gap closes in the DOS. Here we illustrate the PES responses of the system at different temperatures and for different interactions $U$.
\begin{figure}[htb]
\centerline{\includegraphics[width=0.5\textwidth]{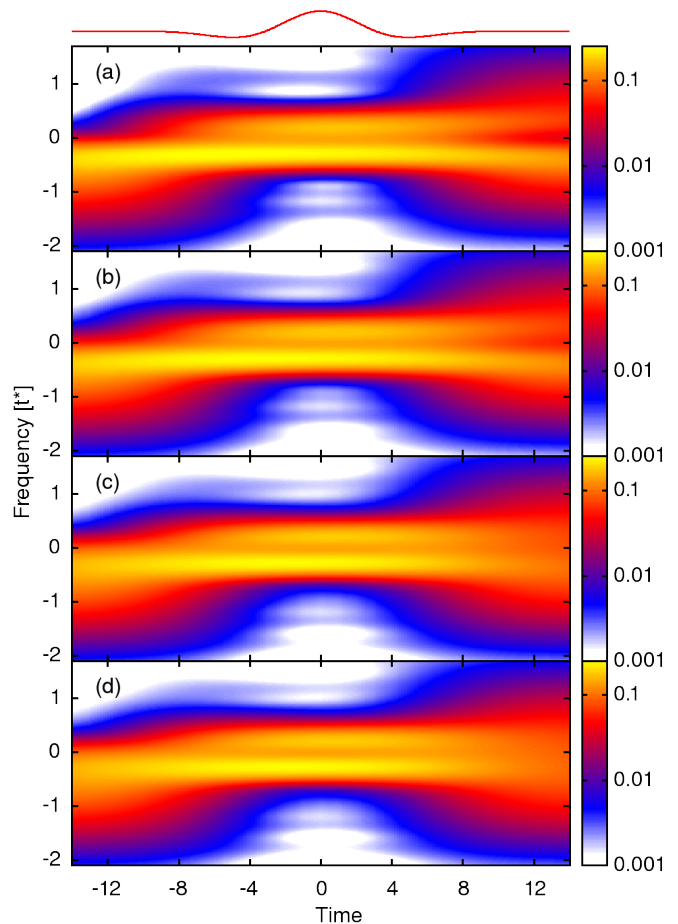}}
 \caption{(Color online.) False color plot of the PES response function for $U=0.5$ at different temperatures with a logarithmic color scale: (a) $T=0.0178$ corresponds to $\Delta n_f=0.49$; (b) $T=0.0278$ 
 corresponds to $\Delta n_f=0.4$; (c) $T=0.0326$ corresponds to $\Delta n_f=0.2$; (d) $T=0.04$ corresponds to $\Delta n_f=0$. The pump field with 
 $E_0=30$ is plotted above, and the probe pulse width is $\sigma_b=7$. Note how the pump pulse does
excite a substantial number of electrons to the upper band, but it also de-excites electrons
(similar to what happens in the simplified CDW model) so that the net excitation at the end of the pulse is small.}
 \label{pes05}
\end{figure}
\begin{figure}[htb]
\centerline{\includegraphics[width=0.5\textwidth]{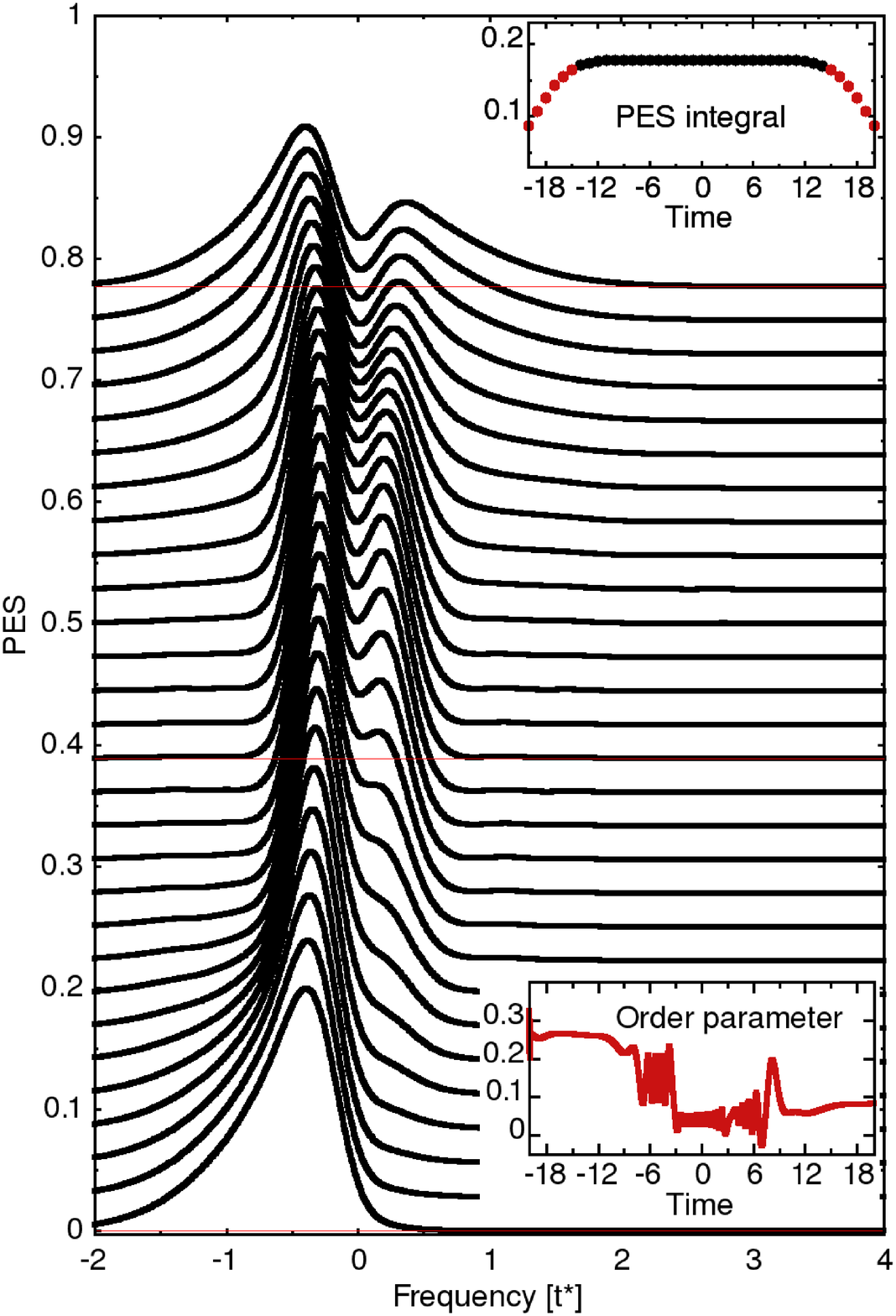}}
 \caption{(Color online.) PES response function for $U=0.5$ at temperature $T=0.0178$ ($\Delta n_f=0.49$); this corresponds to vertical cuts through the false color image in the previous figure. Different curves correspond to different
 time delays $t_0{'}$ for the probe pulse and have been offset in the vertical for clarity. Thin red lines are a guide to the eye. The upper inset shows results for the total spectral weight of the PES response for different $t_0{'}$. The loss of weight at the edges signifies that there is not enough information in the calculated Green's function to properly construct the PES. Data from those red points are not included in the main figure. The lower inset
shows the time evolution of the conduction electron order parameter, which goes from high order to low order after the pulse has ended.}
 \label{pes05wf}
\end{figure}

In Fig.~\ref{pes05}, we plot the PES response function at different probe pulse times $t_0{'}$ in Eq.~(\ref{probe}) for an interaction strength $U=0.5$ and for different 
temperatures. In the normal state, this case corresponds to a correlated metal and the equilibrium DOS for different temperatures is shown in Fig.~\ref{dosu05}. At the top of Fig.~\ref{pes05}, we show the pump pulse which is the electric field from Eq. (\ref{field}) with $E_0=30$.
The width of a probe pulse is equal to $\sigma_b=7$ (the same value will be used for all PES results presented here). We check the 
accuracy of the results with the sum rules in Eq.~(\ref{sumrules}). In our calculations, we discretize a real time interval $t\in[-20,20]$ with three values of 
$\Delta t=0.066,~0.05, \text{and}~0.033$ and then we quadratically extrapolate the result to the $\Delta t=0$ case. We have found that in this case the results accurately satisfy  the sum rules only for large amplitudes  of the field ($E_0\gtrsim20$). To obtain accurate results for smaller fields rapidly becomes problematic due to the small discretization size required and the increase in the size of the matrices used in the calculation.

Figure~\ref{pes05}(a) corresponds to the lowest temperature $T=0.0178$, when the order parameter is equal to $\Delta n_f=0.49$. In Fig.~\ref{pes05wf}, we show 
an alternative view of this case corresponding to vertical cuts through the data. We plot the PES for  times $t_{0}{'}$ starting from $t_{0}{'}=-14$ and ending at $t_{0}{'}=14$ (even though the range in time for the simulation lies in the
interval $[-20,20]$). This is dictated by a loss in conservation of the total spectral weight of the PES signal at extreme times (caused by a shrinking of the available range of relative times). 

Inset in Fig.~\ref{pes05wf}, we plot the integral of the PES depending on the delay time $t_{0}{'}$, where we depict the region that satisfies the sum rule in black. 
For the times before $t=-10$, there is  PES signal only from electrons in the occupied lower band. 
Then, the applied external field excites electrons to the upper band and also closes the CDW gap. Nevertheless, there are still two clear peaks at $-U/2$ and at $U/2$ and one can see that the gap reforms  after the pump pulse (approx. at $t=10$). During the pulse, there is a significant dressing of the bands
which causes the overall effective bandwidth to shrink (and corresponding peaks to grow and sharpen).
This also results in a small reduction of the spectral gap in the PES signal, which can be seen by the 
inward curving of the leftmost peak, which then curves back as the pulse ends.    One can clearly see the 
band narrowing and spectral gap reduction in the false color plots.

At higher temperatures [Figs.~\ref{pes05}(b)-(d)] the scenario remains similar. A slight difference is seen in Figs.~\ref{pes05}(c)-(d), which corresponds
to the temperatures $T=0.0326$ (with the order parameter $\Delta n_f=0.2$), and $T=0.04$ (with the order parameter $\Delta n_f=0$), respectively. In these cases,
we see that gap completely disappears after the pump pulse, since there is no gap in either equilibrium DOS (see Fig.~\ref{dosu05}). Also, there are no apparent shifts of the peaks of the 
PES signal for these cases, although the band narrowing and sharpening of peaks can be easily seen in the
false-color plots.       

Figure \ref{pes05wf} also shows the conduction electron order parameter in the lower right inset. One
can see that it starts off reasonably flat, then oscillates and is reduced, ending nearly at zero. The accuracy
of these calculatons is on the order of a few percent. Even though the order parameter is sharply reduced, the spectral gap feature remains, primarily because the heavy electron order parameter is
unchanged by the pump.

In the supplemental material\cite{supplemental}, movies that show vertical cuts through the false color images, or which animate the waterfall images, are available. These movies clearly show the sharpening of the peaks, the band narrowing, and the fill in of the gap.

\begin{figure}[htb] 
 \centerline{\includegraphics[width=0.5\textwidth]{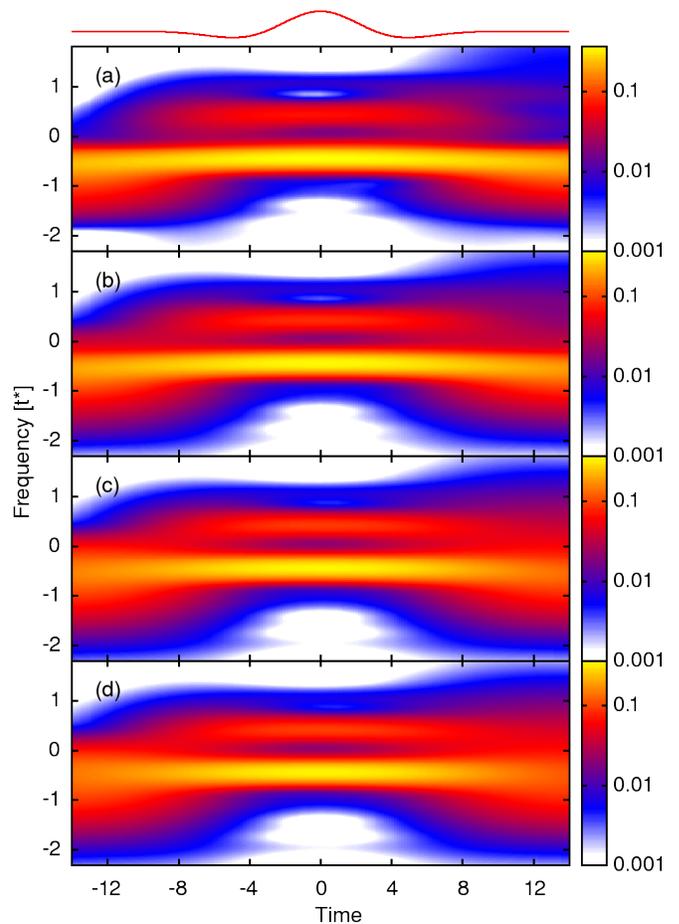}}
 \caption{(Color online.) False color plot of the PES response function for $U=0.86$ (quantum critical material) at different temperatures on a logarithmic color scale: (a) $T=0.03$ corresponds to $\Delta n_f=0.495$; (b) $T=0.047$ 
 corresponds to $\Delta n_f=0.4$; (c) $T=0.053$ corresponds to $\Delta n_f=0.2$; (d) $T=0.06$ corresponds to $\Delta n_f=0$. The pump field with 
 $E_0=30$ is plotted above, and the probe pulse width is $\sigma_b=7$.}
 \label{pes086}
\end{figure}
\begin{figure}[htb]
\centerline{\includegraphics[width=0.5\textwidth]{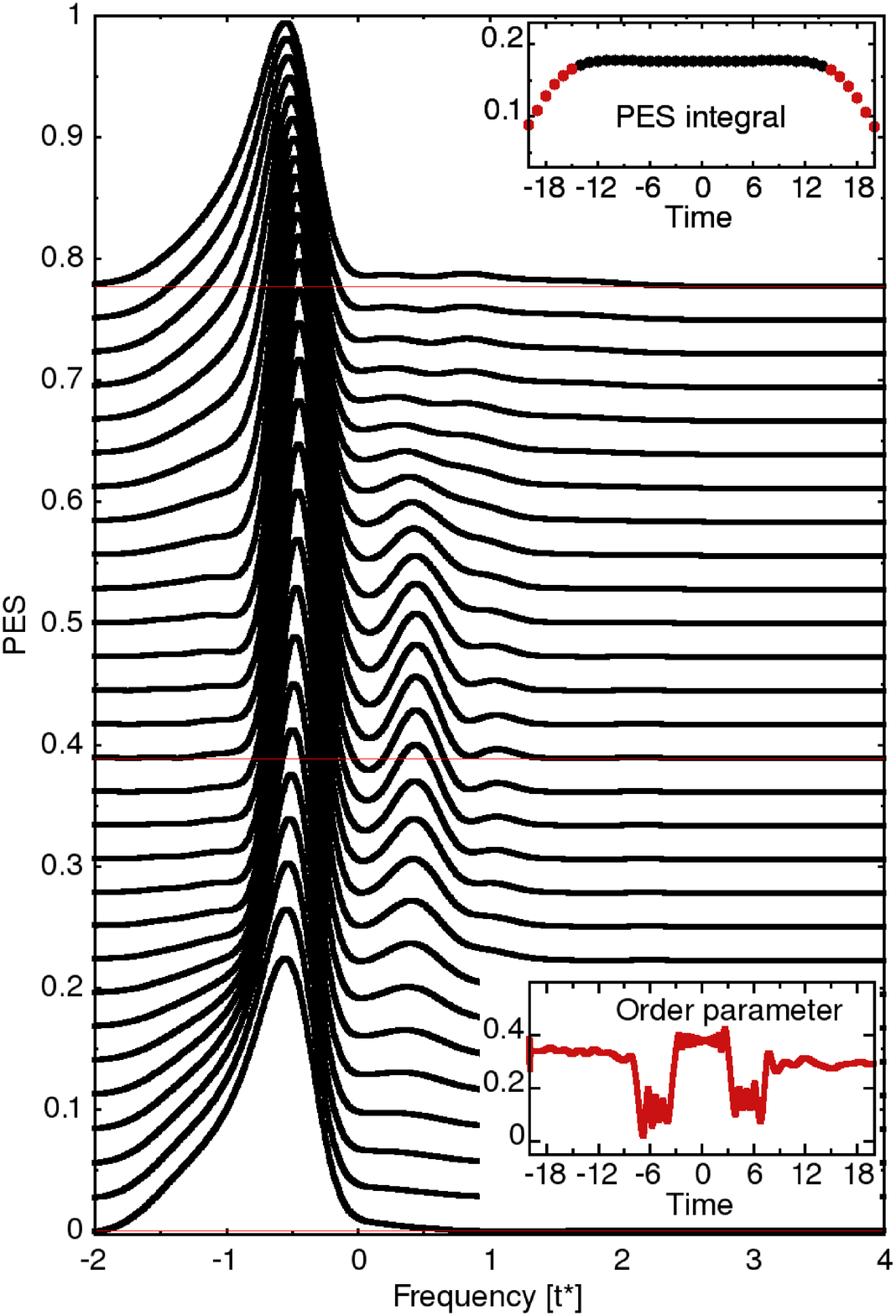}}
 \caption{(Color online.) PES response function for $U=0.86$ at temperature $T=0.03$ ($\Delta n_f=0.495$). Different curves correspond to different
 times $t_0{'}$ of the probe pulse. Thin red lines are a guide to the eye. The upper inset shows results of integration of the PES response for different $t_0{'}$. The lower inset
shows the time evolution of the conduction electron order parameter, which transiently is reduced, but then is restored as the pulse ends. }
 \label{pes086wf}
\end{figure}

In Fig.~\ref{pes086}, we plot the results of the PES response for $U=0.86$, which corresponds to the quantum-critical point. In this case, the equilibrium DOS 
at $T=0$ shows a gap of size $U$. Once $T$ is made nonzero, there is DOS from subgap states that appears starting at the Fermi energy (see Fig.~\ref{dosu086}). In Fig.~\ref{pes086} (a) and Fig.~\ref{pes086wf}, we 
present the PES response function for different times $t_{0}{'}$ for temperature $T=0.03$ when the order parameter is $\Delta n_f=0.495$. In equilibrium, 
the PES response originates from the lower band electrons only. At later times, the pump pulse excites electrons from the lower band into the upper band, providing additional signal. We see 
that the gap disappears at times from $t=-8$ to $t=-4$ then it reforms, then again disappears in the range from $t=4$ to $t=8$, and then it reforms again. 
In contrast to the previous case of $U=0.5$, all the electrons relax into a lower band after the pump pulse: there is not any significant PES response from upper 
band electrons after time $t=10$. This fast relaxation is surprisingly different from the results of the PES response at zero temperature (identical to those of the simplified model).\cite{shen_fr1} In the case of 
simplified model at zero temperature, excited electrons remain in the upper band for long times after the pump pulse because they need a field to de-excite them.\cite{shen_fr1} In the current case, the external electric field excites itinerant electrons, but as they move to the upper band, they also can be de-excited. While the de-excitation must always be less than the excitation, it is this driving of electrons back down to the lower band that dominates the behavior here, leaving the system with few excitations after the pump pulse is gone. In addition, the same spectral bandwidth narrowing, peak sharpening, and spectral
gap reduction seen previously continue to occur transiently when the field is on, and disappear afterwards.

As the temperature is increased the system becomes more and more metallic, which allows for more conventional excitation (see Figs.~\ref{pes086} (c)-(d)). Electrons relax into the states within a ``gap'' region since, the equilibrium DOS shows no gap at these temperatures 
(see Fig.~\ref{dosu086}) and this process of relaxation is slower. The excitations are longer lived as the de-excitation process appears to be suppressed by these thermal fluctuations. Also, 
the reduction of the spectral CDW gap is reduced, but the band narrowing and sharpening only become slightly weaker with temperature [Figs.~\ref{pes086} (c)-(d)]. 

Figure \ref{pes086wf} also shows the conduction electron order parameter, which starts off reasonably flat, then oscillates and then is restored nearly to its starting value.  This suprising evolution is similar to what
is happening with the PES, which ends looking very similar to the way it started with limited excitation. 
It once again shows how this critical CDW state is efficient in de-exciting electrons from the upper to the lower band due to the metallic density of states it has at the chemical potential.

In the supplemental material\cite{supplemental}, movies  are available that show the sharpening of the peaks, the band narrowing, and the fill in of the gap.

\begin{figure}[htb]
\centerline{\includegraphics[width=0.5\textwidth]{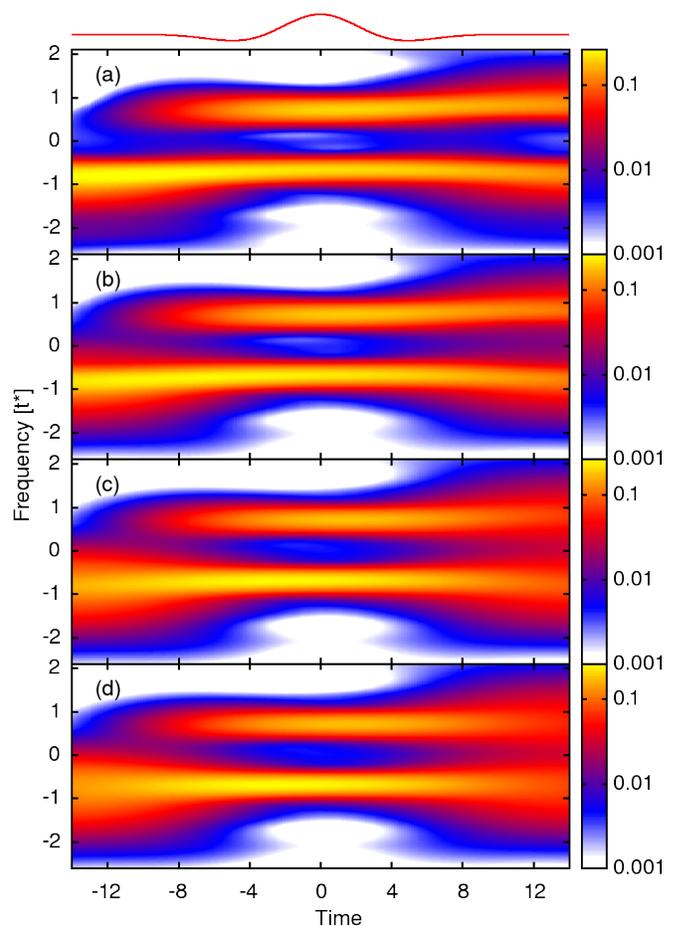}}
 \caption{(Color online.) False color image of the PES response function for $U=1.4$ (critical Mott insulator) at different temperatures on a logarithmic color scale: (a) $T=0.033$ corresponds to $\Delta n_f=0.495$; (b) $T=0.0596$ 
 corresponds to $\Delta n_f=0.4$; (c) $T=0.07$ corresponds to $\Delta n_f=0.2$; (d) $T=0.08$ corresponds to $\Delta n_f=0$. The pump field with 
 $E_0=30$ is plotted above, and the probe pulse width is $\sigma_b=7$.}
 \label{pes14}
\end{figure}
\begin{figure}[htb]
\centerline{\includegraphics[width=0.5\textwidth]{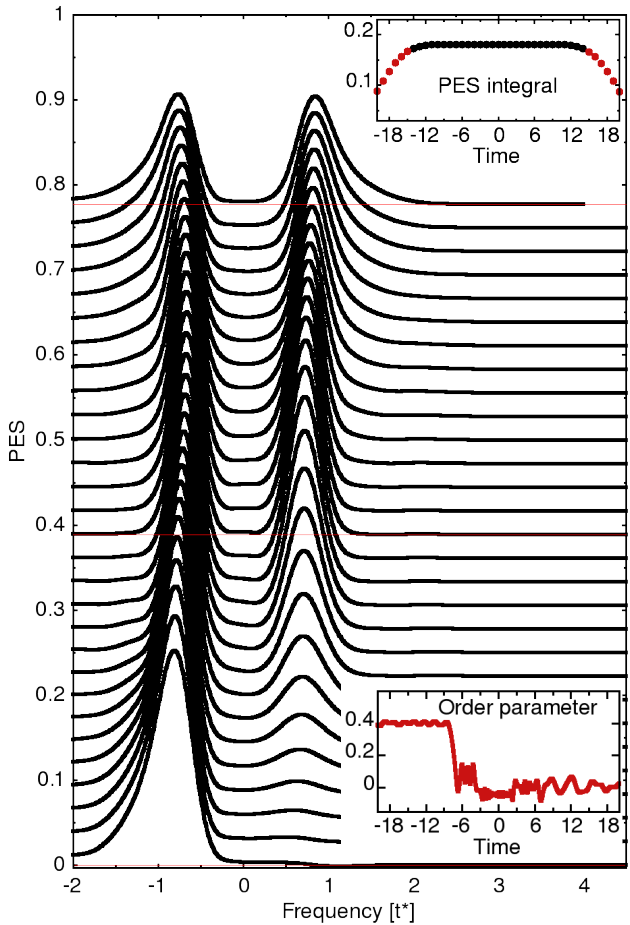}}
 \caption{(Color online.) PES response function for $U=1.4$ at temperature $T=0.033$ ($\Delta n_f=0.495$). Different curves correspond to different
 times $t_0{'}$ of the probe pulse. Thin red lines are a guide to the eye. The upper inset shows the results of the integration of the PES response for different $t_0{'}$. The lower inset
shows the time evolution of the conduction electron order parameter, which goes from high order to low order after the pulse has ended. }
 \label{pes14wf}
\end{figure}

Finally, we present the results of the PES response function for $U=1.4~(\approx\sqrt{2})$ in Fig.~\ref{pes14}. This is the point where the system undergoes the Mott metal-insulator 
transition in the normal state and it corresponds to a strongly correlated CDW. The equilibrium DOS shows the CDW gap at zero temperature and the critical Mott DOS (where the Mott gap is just starting to form) in the uniform phase (at 
temperatures higher than critical $T>T_c$). In Fig.~\ref{dosu14}, we show the equilibrium DOS for different temperatures for $U=1.4$.  Fig.~\ref{pes14}(a) and 
Fig.~\ref{pes14wf} correspond to the lowest temperature $T=0.033$ with $\Delta n_f=0.495$. When the pump pulse hits the system, we see the PES response from the 
electrons in an upper band starting at time $t=-10$. In contrast to the cases discussed above, here a gap remains active during the entire time interval (there are always a small amount of subgap states that preclude a rigorous gap, but they remain small throughout the evolution). At longer times the 
system does tend toward a steady state, as is also seen in case of simplified model at zero frequency\cite{shen_fr1}. At the same time, we see the same band narrowing, peak sharpening, and spectral gap reduction as before.

Figures~\ref{pes14}(c)-(d) correspond to higher temperatures $T=0.7$ ($\Delta n_f=0.2$) and $T=0.8$ ($\Delta n_f=0$), respectively. In these cases, the gap rapidly disappears 
at times from $t=-11$ to $t=-9$, then reforms for the period from $t=-8$ to $t=8$ (during the pump pulse), and then disappears again at 
longer times. This is explained by the fact that there are significant subgap states within the gap (at these temperatures) which accelerates the process of de-excitation.  

Figure \ref{pes14wf} also shows the order parameter, which acts more like the generic case. It starts
off flat, oscillates and is reduced, nearly to zero. Even so, the spectral gap features remain those of the CDW and not of the Mott phase, because of the heavy electron order. This occurs even when nearly half of the electrons are excited into the upper band, which is the maximum one expects for a system approaching infinite temperature. Surprisingly, this Mott phase has more total excitation than the metal or the quantum critical CDW, both which have smaller spectral gaps.

Once again, in the supplemental material\cite{supplemental}, movies  are available that show the sharpening of the peaks, the band narrowing, and the fill in of the gap.

Since the PES spectra is convolved with the probe envelope function, we also show how such
a convolution affects the equilibrium DOS. It removes a number of the sharp structures in the DOS, and
helps explain why those sharp structures are not seen in the theoretical PES response functions that we 
plotted earlier. These plots also show what the PES would be in the quasiequilibrium approximation\cite{pump-probe_theory} (if one also multiplied by the corresponding Fermi-Dirac distribution function). Namely, we calculate
the convolved DOS of the equilibrium system with the pump-probe pulse as follows
\begin{equation}
 P(\omega)=-i\int d\nu G^{r}(\omega-\nu)|\tilde{s}(\nu)|^{2}/2\pi,
 \label{conv}
\end{equation}
where $\tilde{s}(\nu)$ is the Fourier transformation of $s(t)$ in Eq.~(\ref{probe}), and $G^{r}(\omega)$ is total (sum of the $A$ and $B$ sublattices) equilibrium retarded 
Green's function.   

\begin{figure}[htb]
\centerline{\includegraphics[width=0.5\textwidth]{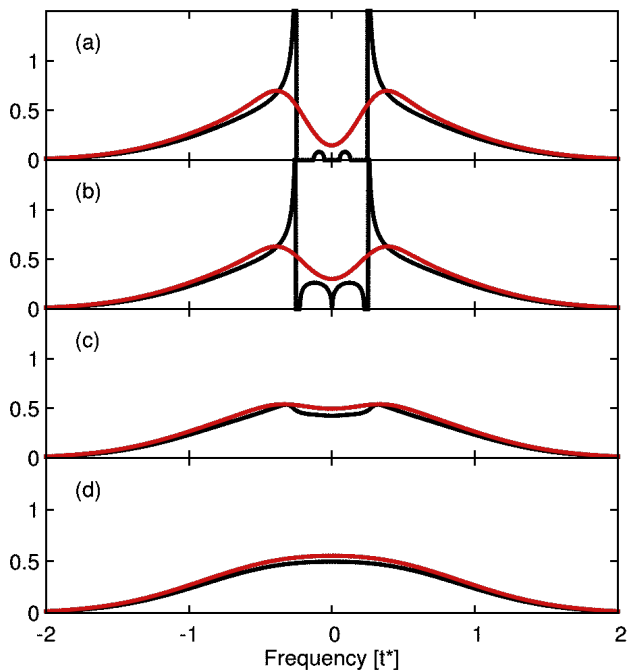}}
 \caption{(Color online.) DOS (black) and convolved DOS (red) for $U=0.5$ at different temperatures: (a) $T=0.0178$ corresponds to $\Delta n_f=0.49$; (b) $T=0.0278$ 
 corresponds to $\Delta n_f=0.4$; (c) $T=0.0326$ corresponds to $\Delta n_f=0.2$; (d) $T=0.04$ corresponds to $\Delta n_f=0$.}
 \label{conv05}
\end{figure}

In Fig.~\ref{conv05}, we compare the convolution from Eq.~(\ref{conv}) to the equilibrium DOS for $U=0.5$. The panels from (a) to (d) correspond to different temperatures 
starting from the lowest $T=0.0178$ to the highest $T=0.04$, respectively. At high temperatures the convolution and the DOS are almost the same in Figs.~\ref{conv05}(c)-(d),
implying that the form of the probe pulse does not play a significant role here. But, at low temperatures in Figs.~\ref{conv05}(a)-(b) the convolution does not show a complete gap and 
does not distinguish separate subgap states. As was mentioned above, one may improve the energy resolution by increasing the width of the pulse, but this decreases time
resolution\cite{spie}.

\begin{figure}[htb] 
 \centerline{\includegraphics[width=0.5\textwidth]{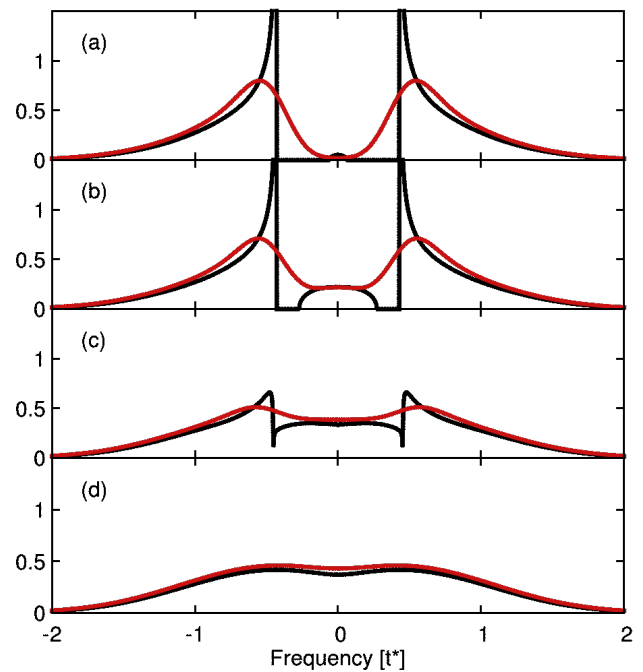}}
 \caption{(Color online.) DOS (black) versus convolved DOS (red) for $U=0.86$ at different temperatures: (a) $T=0.03$ corresponds to $\Delta n_f=0.495$; (b) $T=0.047$ 
 corresponds to $\Delta n_f=0.4$; (c) $T=0.053$ corresponds to $\Delta n_f=0.2$; (d) $T=0.06$ corresponds to $\Delta n_f=0$.}
 \label{conv086}
\end{figure}

Similarly, in Fig.~\ref{conv086}, we present the results for the convolution for quantum critical case of $U=0.86$. In this case, the most interesting result is that we see a gap at the lowest temperature
at zero frequency [Fig.~\ref{conv086}(a)] while the equilibrium DOS shows states at zero frequency. This explains how a ``gap'' in the PES response persists for whole time
interval in Fig.~\ref{pes086}(a) and Fig.~\ref{pes086wf}. 

\begin{figure}[htb] 
 \centerline{\includegraphics[width=0.5\textwidth]{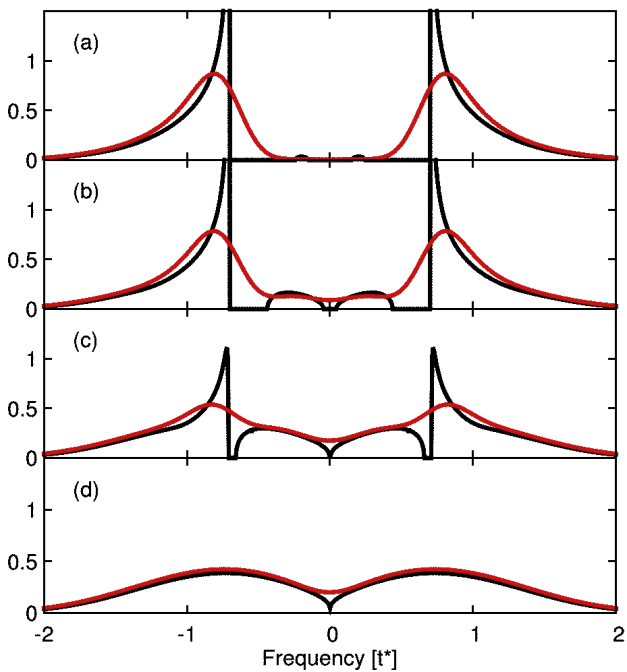}}
 \caption{(Color online.) DOS (black) versus convolved (red) for $U=1.4$ at different temperatures: (a) $T=0.033$ corresponds to $\Delta n_f=0.495$; (b) $T=0.0596$ 
 corresponds to $\Delta n_f=0.4$; (c) $T=0.07$ corresponds to $\Delta n_f=0.2$; (d) $T=0.08$ corresponds to $\Delta n_f=0$.}
 \label{conv14}
\end{figure}

Figure~\ref{conv14} shows the results for the convolution for $U=1.4$. Again, we find that there is a complete gap in the convolution at the lowest temperature with 
no signs of subgap states. At the highest temperature [in Fig.~\ref{conv14}(d)], the convolution is similar to the equilibrium DOS, with a clear two-hump structure.

\section{Conclusions}
\label{sec:3}

In this work, we presented our results on the time-resolved pump-probe PES response function in the CDW ordered phase of the Falicov-Kimball model. We described 
the general formalism to solve for the two-time Green's function defined on the Kadanoff-Baym-Keldysh contour within the nonequilibrium DMFT in the CDW ordered  
phase. These results are numerically exact, but we were forced to restrict ourselves to the time interval $t\in[-20,20]$, and to the
case of a large electric field amplitude for the pump pulse. Similar calculations for small fields require significantly more computer time. We examined three cases 
with different Coulomb interactions, which correspond to the correlated metal, the quantum-critical point for the CDW, and the critical point for the Mott insulator in normal state. Further, we 
examined different temperatures, starting from close to zero temperature (when the system is fully ordered), to a temperature above $T_c$, when the system
is in the normal state. We have also analyzed the role of the form of the probe pulse by comparing the convolved equilibrium DOS with the equilibrium DOS. 

The main counter-intuitive result that we found is that including many-body correlations into the CDW
greatly enhances the relaxation and de-excitation of the system. In particular, the quantum critical case is quite difficult to excite by the pump pulse. The most likely explanation for why this occurs is that the
excitation and de-excitation processes are nearly balanced in this case, making it hard to generate
net electron transfer from the lower to the upper band. {\it Surprisingly, it is much easier to excite a Mott
insulator than it is the quantum-critical CDW or a weakly coupled CDW}. We do not have a full understanding as to why and how
the quantum correlations conspire to remove the excitations in this system; this occurs in the absence of a thermal bath that the system is coupled to. 

The other interesting result is that there is a significant spectral CDW gap shrinkage, bandwidth narrowing, and peak sharpening in the field-dressed PES, which becomes less pronounced as temperature rises. 
Some of these features could be related to recent experiments~\cite{schmitt2}, where a partial closing of the spectral CDW gap is seen, but not a full closing.
Our theory is general and can be applied to other Hubbard-like models. We hope, our results might be helpful for experimentalists to examine real materials which
demonstrate CDW order.

\begin{acknowledgments}
This work was supported by the Department of Energy, Office of Basic Energy Sciences, Division of Materials Sciences and Engineering under Contract Nos. DE-AC02-76SF00515 
(Stanford/SIMES) and DE-FG02-08ER46542 (Georgetown). Computational resources were provided by the National Energy Research Scientific 
Computing Center supported by the Department of Energy, Office of Science, under Contract No. DE-AC02-05CH11231. J.K.F. was also supported by the McDevitt Bequest at Georgetown. 
\end{acknowledgments}

\end{document}